# Einthoven's Triangle Revisited:
# A Mathematical Proof

Pei Jun Zhao, MD MPH
University of Western Ontario, Canada
pzhao26@uwo.ca

**Abstract**

Willem Einthoven is widely considered the father of the electrocardiogram (ECG). In 1912, he proposed a method of determining the electric axis of the heart by using an imaginary equilateral triangle connecting the limb leads, now known as Einthoven's triangle[1,2]. In 1924, Einthoven was awarded the Nobel Prize in Physiology or Medicine for his discovery of the mechanisms of the electrocardiogram[3].

More than a century later, Einthoven's triangle is still at the heart of ECG interpretation. It defines the axes of the ECG leads in the frontal plane, that in turn, determines the axis of the cardiac electric dipole. The method is ubiquitously taught in lectures and applied in clinical settings[4]. But Einthoven did not provide a proof for choosing the equilateral triangle. Future medical literature have not explored its origins.

Thus, 110 years after Einthoven's conception of his famous triangle, this paper provides a formal proof of its derivation to complete this important chapter in medical history and medical education. In addition, the proof determines the geometric conditions for alternative systems of bipolar ECG lead configurations in the frontal plane.

***

To begin, we will set the groundwork by proving Einthoven's law.

**Lemma – Einthoven's Law**

"From the nature of the case there must be a connexion between the curves obtained by the three different leads from the same person. If two forms are known, the third may be calculated from them. The difference between the electrical tensions of leads I and II must be equal to the electrical tension of lead III. This may be formulated : lead II – lead I = lead III."[1]

Proof

In a triangle connecting the three limb leads, let the vertices of the triangle be the limb lead placements – left arm (L), right arm (R), and left foot (F). The direction of limb lead vectors I, II, and III are along the sides of the triangle, RL, RF, and LF, respectively (Figure 1).

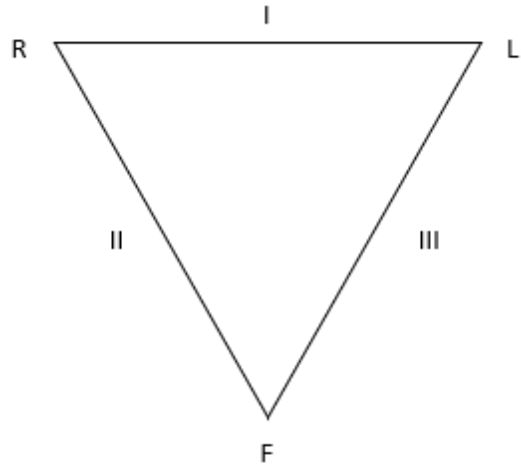

Figure 1: A triangle with the three ECG limb leads as its vertices.

Let $\Phi_{location}$ represent the electric potential at a given body surface location, and $V_{lead}$ represent the voltage of a given ECG lead. Then, by definition, assuming that local perturbations in body surface currents are negligible, the bipolar limb lead voltages are:

$$V_I = \Phi_L - \Phi_R$$
$$V_{II} = \Phi_F - \Phi_R$$
$$V_{III} = \Phi_F - \Phi_L.$$

In electromagnetism, Kirchhoff's voltage law states that the sum of the changes in potential encountered in a complete traversal of any loop of a circuit equals zero. Triangle RLF forms a closed loop. Thus,

$$\Delta V_{LR} + \Delta V_{FL} + \Delta V_{RF} = 0.$$

Therefore,

$$\Delta V_{LR} + \Delta V_{FL} - \Delta V_{FR} = 0.$$
$$\Delta V_{FR} - \Delta V_{LR} = \Delta V_{FL}.$$

That is,

$$(\Phi_F - \Phi_R) - (\Phi_L - \Phi_R) = (\Phi_F - \Phi_L)$$
$$V_{II} - V_I = V_{III.}$$

Please note that all bipolar ECG lead configurations must satisfy Einthoven's Law. This law can be verified on any ECG tracing.

***

Having demonstrated Einthoven's law, we will proceed to derive Einthoven's triangle. With respect to history, we will use mathematical methods known to Einthoven. Although his publications did not reveal a particular mathematical style, one of his contemporary researchers remarked that "It will be evident to the student of mathematics that Einthoven and his followers have used what is known as the polar system of coordinates."[5] Therefore, the same approach will be taken by this paper. The only difference is that while Einthoven measured positive ECG angles going clockwise on the plane, we will preserve the mathematical convention of measuring positive angles counterclockwise.

**Theorem – Einthoven's Triangle**

"The difficulties are solved at once, however, if we apply a schema in which the human body is represented by a plane, homogeneous plate in the form of an equilateral triangle, RLF."[1]

Proof

In the frontal plane, let the unit vector along the three bipolar limb leads be **i**, **j**, **k**, with polar angles of α, β, γ, respectively. Without loss of generality, let the cardiac electric dipole be **p**, with magnitude of 1 and polar angle of θ (Figure 2). The magnitude of the three limb lead tracings are the projections of **p** onto **i**, **j**, and **k**.

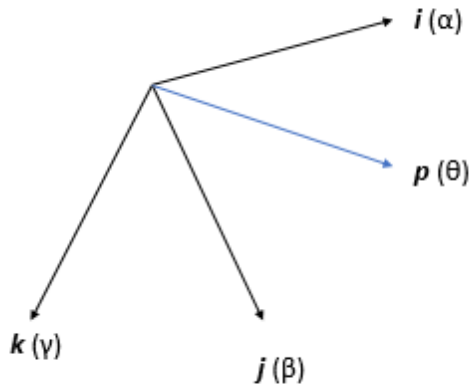

Figure 2: The cardiac dipole and three unit vectors in the direction of the bipolar ECG limb leads.

Einthoven's law states that $\mathbf{p} \cdot \mathbf{i} + \mathbf{p} \cdot \mathbf{k} = \mathbf{p} \cdot \mathbf{j}$. Then,

$$\cos(\theta - \alpha) + \cos(\theta - \gamma) = \cos(\theta - \beta)$$

Expanding the expression,

$$(\cos\theta\cos\alpha + \sin\theta\sin\alpha) + (\cos\theta\cos\gamma + \sin\theta\sin\gamma) = \cos\theta\cos\beta + \sin\theta\sin\beta$$

Collecting terms,
$$\sin\theta(\sin\alpha + \sin\gamma - \sin\beta) + \cos\theta(\cos\alpha + \cos\gamma - \cos\beta) = 0$$

The above equality may be considered in four cases.

**Case 1**: $\sin\alpha + \sin\gamma - \sin\beta = 0$ and $\cos\alpha + \cos\gamma - \cos\beta = 0$.

Rearranging,
$$\sin\alpha + \sin\gamma = \sin\beta \text{ and } \cos\alpha + \cos\gamma = \cos\beta$$

Squaring and adding,
$$(\sin\alpha + \sin\gamma)^2 = \sin^2\beta$$
$$(\cos\alpha + \cos\gamma)^2 = \cos^2\beta$$
$$(\sin\alpha + \sin\gamma)^2 + (\cos\alpha + \cos\gamma)^2 = \sin^2\beta + \cos^2\beta = 1.$$

Expand and simplify,
$$\sin^2\alpha + 2\sin\alpha\sin\gamma + \sin^2\gamma + \cos^2\alpha + 2\cos\alpha\cos\gamma + \cos^2\gamma = 1$$
$$2\sin\alpha\sin\gamma + 2\cos\alpha\cos\gamma + 2 = 1$$
$$\sin\alpha\sin\gamma + \cos\alpha\cos\gamma = -0.5$$

By cosine identity,
$$\cos(\alpha - \gamma) = -0.5.$$

Let the angle between **i** and **k** be between 0 and π, then $\alpha - \gamma = \cos^{-1}(-0.5) = 2\pi/3 = 120°$. In other words, **i** and **k** are 120° apart.

Without loss of generality, suppose $\alpha = 0$, then $\gamma = -2\pi/3 = -120°$. From the conditions in this case,

$$\sin\beta = \sin\alpha + \sin\gamma = \sin 0 + \sin(-2\pi/3) = -\sqrt{3}/2$$
$$\cos\beta = \cos\alpha + \cos\gamma = \cos 0 + \cos(-2\pi/3) = 0.5.$$

Therefore, $\beta = -\pi/3 = -60°$.

Then, geometrically, **i** is along lead I, **j** is along lead II, and **k** is along lead III of Einthoven's equilateral triangle (Figure 3). This configuration holds for any cardiac dipole, regardless of θ.

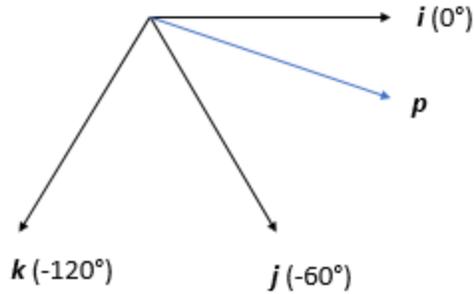

Figure 3: Bipolar ECG limb leads in the configuration of Einthoven's triangle.

**Case 2**: $\sin\theta = 0$ and $\cos\alpha + \cos\gamma = \cos\beta$.

Since $\sin\theta = 0$, let $\theta = 0$, that is, the cardiac dipole has polar angle 0. Then any lead coordinate system that satisfies $\cos\alpha + \cos\gamma = \cos\beta$ also satisfies Einthoven's law.

There are infinite family of solutions of $\alpha$, $\beta$, and $\gamma$ to the equation above. For example, $\alpha = 25°$, $\beta = -56°$, $\gamma = -110°$, and $\theta = 0°$ is a solution (Figure 4). If $\theta$ is non-zero, the entire coordinate system can be rotated by $\theta$, such that $\theta = 0$, and i, j, k have polar angles of $\alpha - \theta$, $\beta - \theta$, and $\gamma - \theta$, respectively.

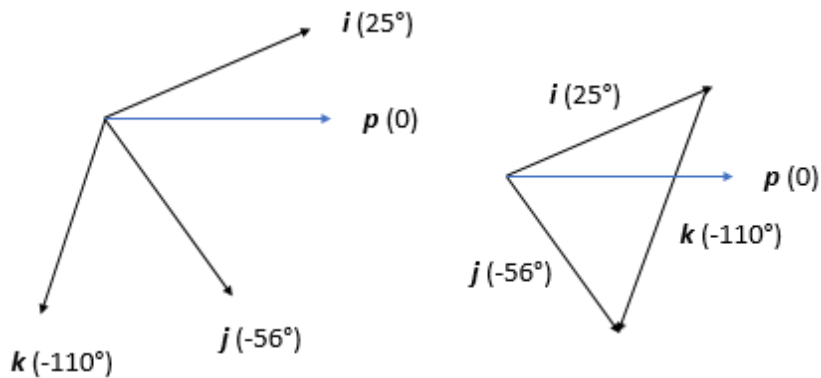

Figure 4: An example of a solution satisfying the conditions in Case 3. On the left is the polar coordinate representation, and of the right the vectors are translated to form a triangle.

Thus, for a specific cardiac dipole, there is an infinite number of possible triangular coordinate systems, and an infinite number of Einthoven's triangles that are not necessarily equilateral.

**Case 3**: $\cos\theta = 0$ and $\sin\alpha + \sin\gamma = \sin\beta$.

Since $\sin\phi = \cos(\pi/2 - \phi) = \cos(\phi - \pi/2)$, Case 3 is Case 2 rotated clockwise by $\pi/2$.

**Case 4**: if $\cos\theta \neq 0$ and $\sin\alpha + \sin\gamma - \sin\beta \neq 0$.

Then rearranging $\sin\theta(\sin\alpha + \sin\gamma - \sin\beta) + \cos\theta(\cos\alpha + \cos\gamma - \cos\beta) = 0$,

We have,
$$\sin\theta(\sin\alpha + \sin\gamma - \sin\beta) = -\cos\theta(\cos\alpha + \cos\gamma - \cos\beta)$$

$$\tan\theta = \frac{\sin\theta}{\cos\theta} = -\frac{\cos\alpha + \cos\gamma - \cos\beta}{\sin\alpha + \sin\gamma - \sin\beta}$$

Therefore, given an ECG lead system with polar angles α, β, and γ, the polar angle θ of the cardiac dipole satisfying Einthoven's law is determined by the equation above.

\*\*\*

**Discussion**

Since Einthoven's publication in 1912 of using an equilateral triangle to determine the electric axis of the heart, his method has been universally adopted and continues to be the standard of ECG interpretation today[4]. Although it has been critically analyzed[6-9] and alternative methods have been proposed such as Burger's triangle[10,11] and Frank's general theory of heart-vector projection[12], none has surpassed the popularity of Einthoven's triangle. But the origin of his triangle has been missing in history[*].

More than 100 years later, this paper presented a formal proof on the derivation of this special triangle using mathematical methods that Einthoven might have used at the time. The proof may be meaningful for both medical history and medical education. Moreover, the equilateral triangle is a special case, as an infinite family of alternative triangles is possible for a given cardiac dipole if certain geometric conditions are satisfied, as detailed in the above proof.

To generalize Einthoven's theory, the frontal plane bipolar ECG lead configuration does not necessarily need to be triangular. Einthoven's triangle can be extended to any closed loop shape, as long as its cardiac dipole projections satisfy Kirchhoff's Law. With increasing use of wearable

---

[*] In medical literature, ascertaining that something does not exist is often more difficult than finding an example of something that does exist. So, if the reader has come across the original derivation of Einthoven's triangle, then please feel free to contact me. Thank you.

Einthoven's Triangle Revisited, 7

ECG devices, such as smart watches, electronic patches, and other sensors with non-conventional lead placements, revisiting the mathematics of ECG lead theory can also help the interpretation of alternative ECG systems.

In reality, the human body is not a homogenous conductor and the electric field of the heart is more complex than a dipole. Despite these simplifications assumed by Einthoven, his model still offers a good approximation for understanding cardiac electrophysiology and for guiding clinical diagnosis. The simplicity, elegance, and practicality of Einthoven's triangle has ensured its enduring status in the medical community.